\documentclass[journal=jpcl,manuscript=article, layout=twocolumn]{achemso}

\usepackage[version=3]{mhchem} 
\usepackage{xcolor} 
\usepackage[capitalise]{cleveref}
\crefname{equation}{reaction}{reactions}
\Crefname{equation}{Reaction}{Reactions}
\usepackage{multirow}
\usepackage[switch, left]{lineno}
\usepackage[colorinlistoftodos,prependcaption,textsize=small]{todonotes}

\author{Germ\'an Molpeceres}
\affiliation[University of Stuttgart]
{Institute for Theoretical Chemistry, University of Stuttgart, 70569, Stuttgart, Germany}
\email{molpeceres@theochem.uni-stuttgart.de}
\author{Johannes K\"astner}
\affiliation[University of Stuttgart]
{Institute for Theoretical Chemistry, University of Stuttgart, 70569, Stuttgart, Germany}
\author{Gleb Fedoseev}
\affiliation[STRW]{Laboratory for Astrophysics, Leiden Observatory, Leiden University, PO Box 9513, 2300 RA Leiden, The Netherlands}
\alsoaffiliation[URFU]{Research Laboratory for Astrochemistry, Ural Federal University, Kuibysheva St. 48, 620026 Ekaterinburg, Russia}
\author{Danna Qasim}
\affiliation[STRW]{Laboratory for Astrophysics, Leiden Observatory, Leiden University, PO Box 9513, 2300 RA Leiden, The Netherlands}
\alsoaffiliation[NASA]{Current address: Astrochemistry Laboratory, NASA Goddard Space Flight Center, Greenbelt, MD 20771, USA}
\author{Richard Sch\"omig}
\affiliation[University of Stuttgart]
{Institute for Theoretical Chemistry, University of Stuttgart,  70569, Stuttgart, Germany}
\author{Harold Linnartz}
\affiliation[STRW]{Laboratory for Astrophysics, Leiden Observatory, Leiden University, PO Box 9513, 2300 RA Leiden, The Netherlands}
\author{Thanja Lamberts}
\affiliation[LIC]{Leiden Institute of Chemistry, Gorlaeus Laboratories, Leiden University, PO Box 9502, 2300 RA Leiden, The Netherlands}
\alsoaffiliation[STRW]{Laboratory for Astrophysics, Leiden Observatory, Leiden University, PO Box 9513, 2300 RA Leiden, The Netherlands}
\email{a.l.m.lamberts@lic.leidenuniv.nl}

\title[C+H2O]
  {Carbon Atom Reactivity with Amorphous Solid Water: \ce{H2O} Catalyzed Formation of \ce{H2CO}}

\abbreviations{IR,NMR,UV}

\begin{document}
\begin{abstract}
We report new computational and experimental evidence of an efficient and astrochemically relevant formation route to formaldehyde (\ce{H2CO}). This simplest carbonylic compound is central to the formation of complex organics in cold interstellar clouds, and is generally regarded to be formed by the hydrogenation of solid-state carbon monoxide. We demonstrate \ce{H2CO} formation \emph{via} the reaction of carbon atoms with amorphous solid water. Crucial to our proposed mechanism is a concerted proton transfer catalyzed by the water hydrogen bonding network. Consequently, the reactions \ce{^3C + H2O -> ^3HCOH} and \ce{^1HCOH -> ^1H2CO} can take place with low or without barriers, contrary to the high-barrier traditional internal hydrogen migration. These low barriers or absence thereof explain the very small kinetic isotope effect in our experiments when comparing the formation of \ce{H2CO} to \ce{D2CO}. Our results reconcile the disagreement found in the literature on the reaction route: \ce{C + H2O -> H2CO}.
\end{abstract}
\begin{center}
\includegraphics[width=2.0in, height=2.0in]{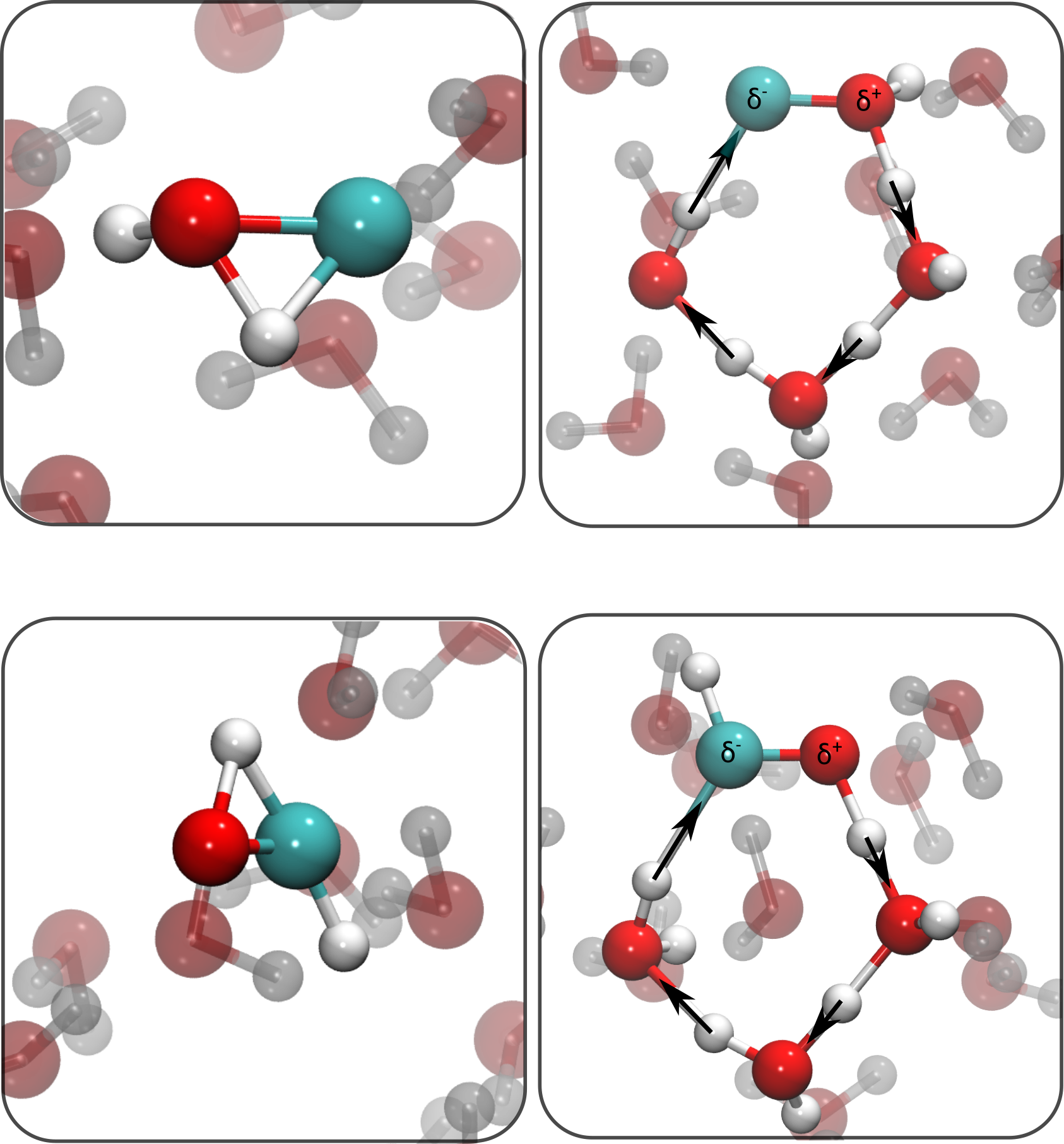} \\
TOC graphic
\end{center}

In molecular clouds where stars are born, temperatures can be as low as 10~K, and water is the main component of the ice mantles coating micron-sized dust grains.\cite{Boogert2015} On the surface of these grains, a rich chemistry accounts for much of the chemical complexity of the known interstellar complex organic molecules (COMs).\cite{Herbst2009,Jin2020,Manigand:2020,Yang:2021} At the core of COM synthesis in space is the formation of bonds to carbon atoms, which, in turn, depends upon the main reservoir of carbon. In the translucent stage of a molecular cloud or under influence of cosmic-ray irradiation, carbon is predominantly present in its atomic form C($^{3}P_{0}$). \cite{Dishoeck:1988, Snow2006,Langer:1976,Keene:1985,Papadopoulos:2004,Burton:2015,Bisbas:2019} The interaction of atomic carbon with water has been extensively studied, both experimentally and theoretically \cite{Ortman1990, Flanagan1992,Schreiner2006, Hickson2016, Li2017, Wakelam2017, Shimonishi2018, Duflot2021}. The particularly puzzling contrast between the gas phase and the condensed phase is illustrated by the reduction of the \ce{C-O} distance of 1.89~{\AA} in the gas-phase \ce{C-H2O} complex,\cite{Schreiner2006} to  1.50~{\AA} on average in water clusters \cite{Shimonishi2018, Duflot2021}. In these clusters, the \ce{C-O} distance depends on the local geometry of the adsorption site, the formation of a \ce{^{3}C-OH2} complex \cite{Shimonishi2018} and/or the formation of a \ce{[COH- + H3O+]} complex \cite{Duflot2021}.

Formaldehyde (\ce{H2CO}) is the lowest energy isomer of C-\ce{H2O} and is a key species in astrochemical reaction networks as a precursor of COMs.\cite{Chuang2016, Butscher2017} At present, chemical models consider the sequential hydrogenation of solid CO as its main surface formation route.\cite{Watanabe2002, Fuchs2009} This links \ce{H2CO} formation to later stages of a molecular cloud, where CO ice is typically abundant.\cite{Caselli1999, Boogert2015}. The formation of \ce{H2CO} may proceed also in the gas phase.\cite{TerwisschavanScheltinga2021}%


The \ce{^{3}C + H2O -> \emph{products}} reaction in the gas phase proceeds through quantum tunneling, indicated by a lower reactivity with \ce{D2O}.\cite{Hickson2016} In matrix-isolation studies, however, no products for the same reaction have been detected.\cite{Ortman1990, Schreiner2006} Furthermore, C atoms generated in an arc discharge react with water at 77~K.\cite{Flanagan1992} A subsequent GC/MS analysis showed a variety of aldoses, and, among them, formaldehyde. Very recent experimental work indicates the formation of \ce{H2CO} in the solid state on water ice.\cite{potapov2021} 


In this letter we provide a detailed explanation of the formation of formaldehyde from the reaction of carbon atoms with amorphous solid water within a theoretical framework, supported by tailored experiments. Our work simulates the earliest stages in the star formation process, before the catastrophic CO freeze-out stage,\cite{Pontoppidan:2003} \emph{i.e.}, before the CO hydrogenation chain dominates \ce{H2CO} formation. Our experiments probe the kinetic isotope effect, comparing the products of the reactions \ce{C + H2O / HDO / D2O}. We resolve the apparent discrepancy between earlier studies by introducing our proposed mechanism in which water molecules collectively catalyze the reaction \emph{via} a proton transfer.
We show that this transfer operates throughout the entire \ce{^{3}C + H2O -> H2CO} reaction sequence. This implies that formaldehyde ice abundances could be higher than previously anticipated, especially in the early stages of star formation, which is soon expected to be probed by James Webb Space Telescope (JWST) observations of interstellar ices.\cite{Keane2001,Boogert2015}


\begin{figure*}
    \centering
        \includegraphics[width=0.65\textwidth]{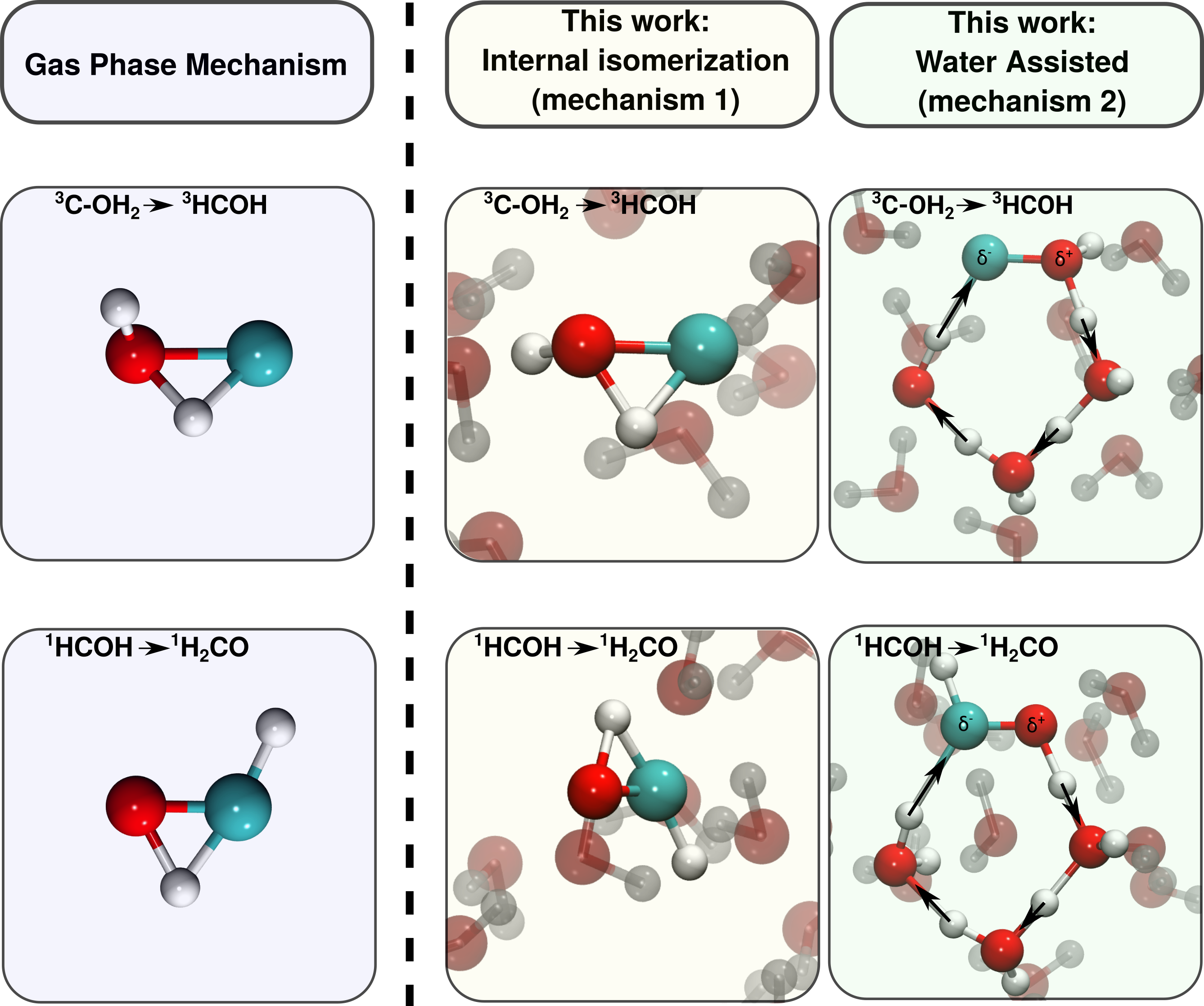}
    \caption{Transition state geometries for \cref{eqn:isom} in the upper row, and \cref{eqn:isom2}, in the lower row, in the gas phase (left column) and for both the traditional internal isomerization mechanism \textbf{1} (middle column) and the concerted water-assisted mechanism \textbf{2} (right column). Atoms directly involved in the reaction mechanism are depicted in full color, water molecules not participating in the reaction are transparent in the background. Color code: Teal -- Carbon, Red -- Oxygen, White -- Hydrogen}
    \label{fig:ts_c}
\end{figure*}

\begin{figure*}
    \centering
        \includegraphics[width=\textwidth]{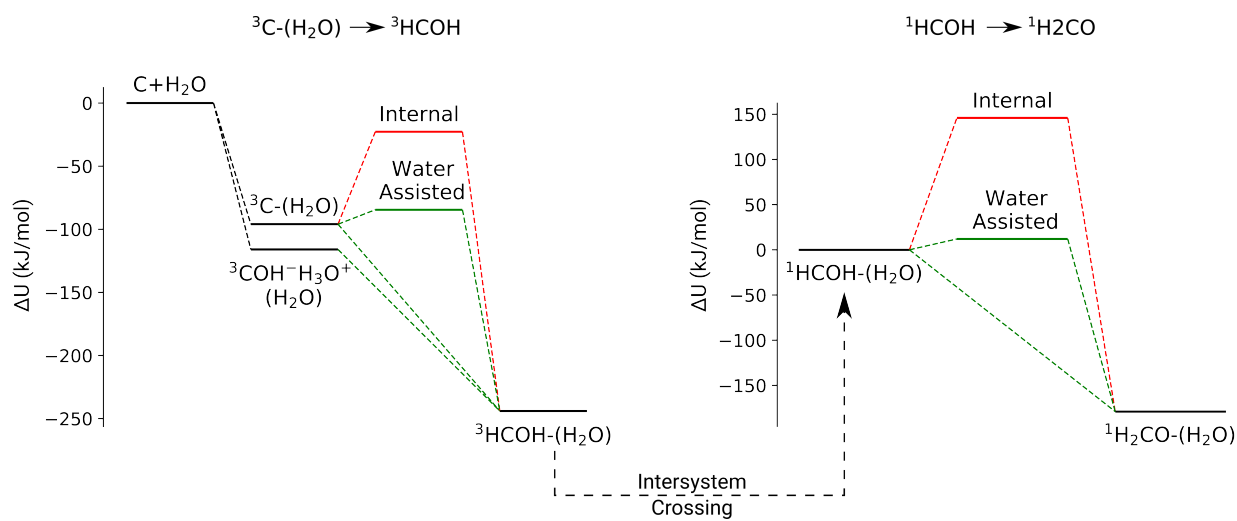} 
    \caption{Schematic reaction profiles for \cref{eqn:isom,eqn:isom2} within the internal isomerization mechanism (Mechanism \textbf{1}, red) and the water assisted mechanism (Mechanism \textbf{2}, green). Note that green dashed lines here represent paths with a small barrier and barrierless paths. The minima are taken as the average binding energies listed in Table \ref{tab:binding} and the saddle points as the highest activation energies in Table \ref{tab:summary}. For \ce{^1H2CO}, the binding energy is taken as the average of the endpoints of intrinsic reaction coordinate calculations. The reader is referred to the text for an extensive discussion of the energetics of the reaction.}
    \label{fig:profile}
\end{figure*}

We modeled the formation of \ce{H2CO} on amorphous water ice clusters considering two distinct mechanisms: \textbf{1}. Traditional internal isomerization, and \textbf{2}. Concerted water-assisted isomerization, and compare this to the internal isomerization in the gas phase. Below, we list the relevant reaction steps for which we wish to point out that in both mechanisms all species are at all times adsorbed on a surface. Thus, mechanism \textbf{1} may resemble the process in the gas phase, but with adsorbed reactants. We present transition states for both mechanisms and the analogs in the gas phase in \cref{fig:ts_c} to guide the reader through the discussion. For both mechanisms the reaction sequence starts with:


\begin{align}
      \ce{^3C + H2O} & \ce{-> ^3C-OH2} \label{eqn:complex} \\
      \ce{^3C-OH2} & \ce{-> ^3HCOH}\;. \label{eqn:isom}
\end{align}

\Cref{eqn:complex} represents the complex formation and \cref{eqn:isom} an insertion or isomerization \emph{via} internal hydrogen migration.


Subsequently, also for both mechanisms, an intersystem crossing (ISC) to the singlet surface needs to take place followed by a hydrogen migration to finally yield formaldehyde, either through 

\begin{align}
    \ce{^{3}HCOH} &\ce{-> ^{3}H2CO} \label{eqn:isom3} \\
    \ce{^3H2CO} &\ce{-> ^1H2CO} \label{eqn:ISC2}
\end{align}
or
\begin{align}
    \ce{^{3}HCOH} &\ce{-> ^{1}HCOH} \label{eqn:ISC}\\
    \ce{^{1}HCOH} &\ce{-> ^{1}H2CO} \;. \label{eqn:isom2} 
\end{align} 
\Cref{eqn:isom,eqn:isom3,eqn:isom2} proceed with high activation barriers in the gas phase.\cite{Schreiner2008, Hickson2016, Li2017} We test here to what extent this holds in mechanism \textbf{1}. Mechanism \textbf{2} follows the same reaction steps, but a surrounding ice water molecule acts as a proton donor (\emph{e.g.}, $\delta^+$) for \cref{eqn:isom,eqn:isom2} (depicted by the arrows in \cref{fig:ts_c}), greatly lowering the reaction barriers, see also \cref{fig:profile}. Note that \cref{eqn:isom3} does not proceed in mechanism \textbf{2} and is not included in either \cref{fig:ts_c} or \cref{fig:profile}.


Reactions are computationally studied individually by placing the respective reactants on a water ice cluster, \ce{(H2O)_{14}}. We randomly placed the adsorbates: \ce{^3C} , \ce{^{3}HCOH}  and \ce{^{1}HCOH} on different pre-optimized \ce{(H2O)14} clusters, constructed \emph{via} molecular dynamics heating followed by a geometry optimization using \textit{ab-initio} methods and determining 48, 42 and 44 binding sites, respectively. The ranges of binding energies along with their average value for each adsorbate are gathered in \cref{tab:binding}. From these binding sites both mechanism \textbf{1} and mechanism \textbf{2} were explored, in search of transition states for \cref{eqn:isom,eqn:isom3,eqn:isom2}. We utilized a single shallow binding site for mechanism \textbf{1}, whereas we sampled three binding sites for mechanism \textbf{2}, because the latter is more dependent on the binding site geometry. The level of theory used is MPWB1K/def2-TZVP on optimized structures at the MPWB1K/def2-SVP level, and benchmarked against gas-phase values using the UCCSD(T)-F12a/cc-pVTZ-F12//MPWB1K/def2-TZVP level of theory (also included in Table \ref{tab:summary}). This yields very good results within the variability provided by the different binding sites. Note that our values in the triplet channel are also in good agreement with the results of \citeauthor{Hickson2016}\cite{Hickson2016} and \citeauthor{Li2017}\cite{Li2017}


To describe the striking difference between both mechanisms, that ultimately renders a completely different physicochemical picture, we first focus on the reaction sequence in mechanism \textbf{1} followed by mechanism \textbf{2}. \Cref{fig:profile} gives a schematic representation of the energy profile of both mechanisms. \cref{tab:binding} and \ref{tab:summary} build on \cref{fig:profile}, providing detailed values of the binding energies and activation barriers. All binding energies are computed from the difference between the energy of the adsorbate+cluster minus the sum of the separate components while activation energies are obtained relative to the reactant on the surface obtained from intrinsic reaction coordinate (IRC) calculations. All binding energies (\cref{tab:binding}) and activation energies (\cref{tab:summary}) are reported including zero-point energy (ZPE). More details on the protocol followed for the \textit{ab-initio} calculations can be found in the Supporting Information.


\begin{table*}[h]
    \centering
        \caption{Binding energy ranges and averages for the different species and complexes on water clusters considered in this work. Average binding energies are obtained as the mean of the individual binding energies in each binding site. Please note that the binding energies provided here apply to both mechanisms \textbf{1} and \textbf{2}.}
    \label{tab:binding}
    \resizebox{\textwidth}{!}{
    \begin{tabular}{ll|cc}
\hline
       \multicolumn{2}{l|}{Adsorbate - cluster system} & Binding energy range  & Average binding energy  \\
                    &                                 & (kJ/mol)              & (kJ/mol) \\
    \hline \hline
             & \ce{-> ^3C - (H2O)_14}$^{i}$  & 60 - 133 & 96  \\
\ce{^3C + (H2O)_14}$^{ii}$ & \ce{-> ^3[COH- H3O+] - (H2O)_13} & 88 - 143  & 116 \\
             & \ce{-> ^3HCOH - (H2O)_13}  & 231 - 268  & 244 \\
\ce{^3HCOH + (H2O)_14} & \ce{-> ^3HCOH - (H2O)_14} & 7.5 - 52.5  & 29.4  \\
\ce{^1HCOH + (H2O)_14} & \ce{-> ^1HCOH - (H2O)_14} & 7.8 - 95.6 & 52.4  \\
\hline
\multicolumn{4}{l}{$^{i}$\footnotesize{a) The values provided in the literature for \ce{^3C} on ASW (average binding) are 116 kJ/mol and 79 kJ/mol.\cite{Shimonishi2018, Duflot2021} }} \\
\multicolumn{4}{l}{\footnotesize{b) \ce{^3C} adsorption on ice I$_{h}$ reported binding energies are 153 and 127 kJ/mol depending on the binding site.\cite{Ferrero2020}}} \\
\multicolumn{4}{l}{$^{ii}$\footnotesize{According to our sampling: 71\% \ce{^3C - (H2O)_14}; 19\% \ce{^3[COH- H3O+] - (H2O)_13}; and 10\% \ce{^3HCOH - (H2O)_13}}}

    \end{tabular}
    }
\end{table*}

\begin{table*}
\caption{Activation energies without ($\Delta E_{a}$) and with ZPE ($\Delta U_{a}$) in kJ/mol for \cref{eqn:isom,eqn:isom2,eqn:isom3} in the gas phase, \emph{via} the internal isomerisation mechanism (\textbf{1}) and the water-assisted isomerisation mechanism (\textbf{2}) at the MPWB1K/def2-TZVP//MPWB1K/def2-SVP level of theory. In parentheses UCCSD(T)-F12a/cc-pVTZ-F12//MPWB1K/def2-TZVP reference values. }  
\label{tab:summary} 
\centering  
\begin{tabular}{c | c c | c c  }        
\hline     
Mechanism & Reaction & Binding site $^{i}$ & $\Delta E_{a}$ & $\Delta U_{a}$  \\ 
\hline     
\multirow{2}{*}{Gas phase} & \cref{eqn:isom}  & -- & 69.7 (72.5) & 57.3 (60.1)  \\
                            & \cref{eqn:isom3} & -- & 139.6 (139.7) & 126.6 (126.6)  \\
                             & \cref{eqn:isom2} & -- & 151.3 (144.4) & 134.9 (127.9)  \\
\hline
\multirow{3}{*}{Mechanism 1} & \cref{eqn:isom}  & shallow & 86.9 & 73.3  \\
                             & \cref{eqn:isom3} & shallow & 138.3 & 123.9  \\
                             & \cref{eqn:isom2} & shallow & 163.8 & 146.6  \\
\hline
\multirow{4}{*}{Mechanism 2$^{ii}$} & \multirow{2}{*}{\cref{eqn:isom}} & medium / deep & 30.3--36.9 & 9.7--11.5  \\ 
                             &                   & \ce{[COH- + H3O+]} & 17.5 & -0.31  \\
                             & \cref{eqn:isom2} & medium / 2 deep & 12.8--31.7 & -1.0--12.0  \\
                             
\hline       
\multicolumn{5}{l}{$^{i}$\footnotesize{See text}} \\
\multicolumn{5}{l}{$^{ii}$\footnotesize{\Cref{eqn:isom3} does not proceed in Mechanism 2.}} \\

\end{tabular}
\end{table*}

\textit{Mechanism 1.} The onset of the reaction is the binding of a carbon atom atop amorphous solid water (ASW). Three regimes can be clearly distinguished, see also \cref{tab:binding}. The first regime, \ce{^3C + (H2O)_14 -> ^3C - (H2O)_14} is the one relevant for mechanism \textbf{1}, since there is no spontaneous chemical conversion after interacting with \ce{H2O} ice.\cite{Shimonishi2018, Duflot2021} Within this regime we expect other reactions with carbon to be possible, \emph{e.g}, to form \ce{CH4} after hydrogenation.\cite{Qasim2020}
In order to form \ce{^3HCOH} via internal isomerization, the hydrogen atom of \ce{C-OH2} that migrates should not take part in any hydrogen bonds with the backbone of the cluster, see middle top panel in \cref{fig:ts_c}. Such a site ($U_\text{bind}= 72.8$ kJ/mol) rendered an activation energy of $\Delta U_{a}$= 73.3 kJ/mol.
After formation of \ce{^3HCOH}, chemical conversion to \ce{^3H2CO} is studied through a transition state found for a weak binding site ($U_\text{bind}= 10.3$ kJ/mol) with a resulting barrier of 123.9~kJ/mol. Both activation energies are either higher than or very similar to the corresponding gas-phase reaction,\cite{Li2017} see \cref{tab:summary}. Such barriers are too high to be overcome at 10~K, even on interstellar timescales.


The formation of formaldehyde would require intersystem crossing (ISC) from \ce{^3HCOH} to $^{1}$HCOH \emph{via} \cref{eqn:ISC}. The singlet state is more stable than the triplet one by 88.9 kJ/mol. The explicit simulation of ISC is beyond the scope of the present work, but studies in the gas phase showed that the spin-orbit coupling is strong,\cite{Zhang2009a} which determines the intersystem crossing rate. We expect that the `heavy' O atoms in the ice further enhance spin-orbit coupling. Experimental ISC timescales for \cref{eqn:ISC2} report fast conversion (ns to $\mu$s)\citep{Hwang1999,Kruger2019} and we expect the same for \cref{eqn:ISC}.


Finally, a relatively shallow binding site ($U_\text{bind}= 22$ kJ/mol, with the OH moiety not interacting with the ice) gives an activation energy of $\Delta U_{a}$=146.6 kJ/mol for \cref{eqn:isom2}, \ce{$^{1}$HCOH -> ^1H2CO}, depicted in the middle bottom panel in \cref{fig:ts_c}. The barrier is significantly higher than in the gas phase. We attribute this to a stabilizing effect of the surface on the reactant state, again rendering this pathway unlikely to be relevant at low temperatures. 


In short, as a result of the high barriers involved, also on the surface, the traditional internal isomerization mechanism \textbf{1} is unlikely to lead to formaldehyde on water ices.


\textit{Mechanism 2.} During the adsorption of \ce{^3C} on ASW, not only the \ce{^3C-H2O} structure is found, but also two other outcomes are observed: a \ce{[COH- + H3O+]} complex\cite{Duflot2021} and a direct and barrierless  to \ce{^{3}HCOH}, also hinted at by \citeauthor{Shimonishi2018}\cite{Shimonishi2018} Closer inspection of our optimization trajectories shows that the migrating H is transferred from a polarized water molecule close to the \ce{^{3}C-OH2} adduct, resembling an acid-base process.  Whether or not this chemical conversion takes place thus greatly depends on the local geometry of the water into which the carbon inserts. The Supporting Information details that the barrierless pathway remains barrierless, also for other exchange and correlation functionals.


In addition to this barrierless pathway, concerted transition states leading to \ce{^3HCOH} formation are explored. We depict one in the top right panel in \cref{fig:ts_c}. We found no transition state for the weakly bound C atom, further suggesting that reactions involving free carbon happen in these binding sites. For the strongly and medium bound atoms, the activation energies are $\Delta U_{a}$=9.7 and 11.5 kJ/mol, respectively, a factor $\sim7$ lower than for the mechanism \textbf{1}, see also \cref{fig:profile}. 


Finally, the transition state found for the \ce{[COH- + H3O+]} complex leads to a negative barrier (\emph{i.e.}, barrierless) upon the inclusion of the ZPE correction. In general, including ZPE corrections reduces the barrier more for the water-assisted mechanism, see \cref{tab:summary}, which can be rationalized by the imaginary transition mode involving collective high-frequency water stretching modes. 


Since \ce{^3HCOH} can thus readily be formed, we also examined the conversion to \ce{^3H2CO} through the water-assisted mechanism. However, all three binding sites lead to endothermic reaction paths and thus \cref{eqn:isom3} and, therefore, also \cref{eqn:ISC2} are unlikely to take place. For the intersystem crossing \cref{eqn:ISC}, \ce{^3HCOH -> ^1HCOH}, the same reasoning as mentioned above holds.


Finally, transition states for \cref{eqn:isom2} are found for three different binding sites. Two of them lead to activation energies of $\Delta U_{a}$=9.0 and 12.0 kJ/mol, in agreement with a previous work.\cite{Peters2011} For the third binding site, the reaction is barrierless upon inclusion of ZPE. This is clearly depicted also in the bottom right panel of \cref{fig:ts_c}: the water ice acts as a catalyst for the reaction transferring a proton from the surface hydrogen bond network to the CH moiety. 


Summarizing, the \ce{^{1}HCOH -> ^{1}H2CO} reaction sequence can take place via the water-assisted mechanism. For some binding sites, we find no barrier for the reaction. Combining this finding with the results for \cref{eqn:complex} and \cref{eqn:isom}, opens the possibility of an effectively barrierless reaction pathway for the formation of \ce{H2CO} from the \ce{C + H2O} reaction assuming that intersystem crossing is fast, see \cref{fig:profile}.


\textit{Deuteration and Experiments.} 
An overall low-barrier or barrierless pathway implies that, on ice, a small isotope effect is expected for reactions in which the hydrogen atoms are replaced by deuterium. This is in stark contrast with results obtained in the gas phase finding a significant kinetic isotope effect.\cite{Hickson2016} Furthermore, the so-called deuterium fractionation of molecules is often used as a probe to understand whether formation has taken place in early or late molecular cloud stages in the ISM.\cite{Ceccarelli:2014} The barrierless nature of the reaction could have a consequential effect on formaldehyde deuterium fractionation in interstellar clouds.


The influence of deuterium substitution on the activation energies is evaluated by recomputing barrier heights for three different deuterium substitution cases for the key \cref{eqn:isom,eqn:isom2} in mechanism \textbf{2}: 


\begin{enumerate}
\setlength{\itemsep}{0pt}
\setlength{\parskip}{0pt}
    \item Substitution of the H atom that transfers to the C atom.
    \item Substitution of all H atoms taking part in the water-assisted mechanism (\emph{e.g.}, those in the concerted transition state)
    \item Substitution of all H atoms in the water cluster.
\end{enumerate}
We selected the binding sites associated with the highest and lowest activation energy. Our results are summarized in Table \ref{tab:deuterium}.


\begin{table}[t]
\caption{Activation energies ($\Delta U_{A}$ in kJ/mol) for water assisted \cref{eqn:isom,eqn:isom2} and three deuteration cases, see text, compared to Case 0 without deuteration.}  
\label{tab:deuterium} 
\centering  
\begin{tabular}{c c | c c }        
\hline     
 \multicolumn{4}{c}{Reaction} \\
 \hline     
 \multicolumn{2}{c}{\ce{^{3}C-OH2 -> ^{3}HCOH}} & \multicolumn{2}{c}{\ce{^{1}HCOH -> ^{1}H2CO}} \\    
 \hline
Case & $\Delta U_{A}$  & Case & $\Delta U_{A}$   \\
  \hline
  \multicolumn{4}{c}{Highest Activation Energy} \\
 \hline     
Case 0 & 11.5  & Case 0 & 12.0   \\      
Case 1 & 12.3  & Case 1 & 12.2   \\
Case 2 & 18.3  & Case 2 & 17.4   \\
Case 3 & 17.8  & Case 3 & 17.9   \\
 \hline
  \multicolumn{4}{c}{Lowest Activation Energy} \\
 \hline
Case 0 & -0.3  & Case 0 & -1.0   \\      
Case 1 &  3.4  & Case 1 &  2.3   \\
Case 2 &  5.6  & Case 2 &  3.7   \\
Case 3 &  5.4  & Case 3 &  3.9   \\
  \hline
 \hline
\end{tabular}
\end{table}

The transition states concern a concerted motion, involving many molecules which is nicely illustrated by the fact that the activation energy is only mildly affected in Case 1, but clearly increases for Cases 2 and 3. The increase of the barrier height finds its origin in the delocalized nature of the transition state, significantly reducing the magnitude of the imaginary transition mode. 
Despite the increase, the activation energies remain low enough to expect that a reaction between the carbon atom and deuterated water can take place, for instance \emph{via} tunneling.\cite{potapov2021} Note that the barrierless pathways for \ce{^3HCOH} formation in \cref{eqn:isom} are also still relevant for reactions with HDO or \ce{D2O}.


Our theoretical results were tested against four tailored experiments that we performed following \ce{H2CO} detection in previous works.\cite{Qasim2020,Qasim:2020b,potapov2021}
\cref{tab:exps} outlines the four co-depositions of a mixture of C atoms, \ce{H2O}, and/or \ce{D2O} at 10~K using the SURFRESIDE$^3$ setup.\cite{Qasim:2020b}. 
Experiments 1 and 2 test formation of \ce{H2CO} and \ce{D2CO}. Experiment 3 probes the simultaneous formation of \ce{H2CO}, HDCO and \ce{D2CO} isotopologues from reaction of carbon with \ce{H2O}, HDO, and \ce{D2O}. In particular, HDO is formed as the result of D/H exchange on the metal walls of the dosing line capillaries upon simultaneous injection of \ce{H2O} and \ce{D2O} vapours which thus leads to the deposition of a \ce{H2O}:HDO:\ce{D2O} mixture.
Note that while there is no production of C atoms in our control experiment 4, the source was operated at a low temperature of 1200 $^\circ$C in order to account for the possible effects of contaminants from the degasing C-atom source, such as \ce{H2O} and \ce{CO2}. The growing ice was monitored with Reflection Absorption InfraRed Spectroscopy (RAIRS). More details on the experiments are given in the Supporting Information.


\begin{table}[t]
    \centering
    \caption{Summary of the four key experiments, including the atomic and molecular effective fluxes and the total time of the experiment.}
    \label{tab:exps}
    \begin{tabular}{l|rrrr}
    \hline
    \#     & C & \ce{H2O} & \ce{D2O} & Time \\
        &    \small cm$^{-2}$ s$^{-1}$      & \small cm$^{-2}$ s$^{-1}$      & \small cm$^{-2}$ s$^{-1}$     & \small min \\
        \hline\hline
    1     & $5\times10^{11}$ & $1.2\times10^{13}$ & -- & 50 \\
    2     & $5\times10^{11}$ & -- & $1.4\times10^{13}$  & 50 \\
    3     & $5\times10^{11}$ & $1.4\times10^{13}$ & $1.6\times10^{13}$ & 125 \\
    4     & --         & $1.4\times10^{13}$ & $1.6\times10^{13}$ & 65 \\              

\hline    
    \end{tabular}
\end{table}

\begin{figure*}
    \centering
        \includegraphics[width=0.75\textwidth]{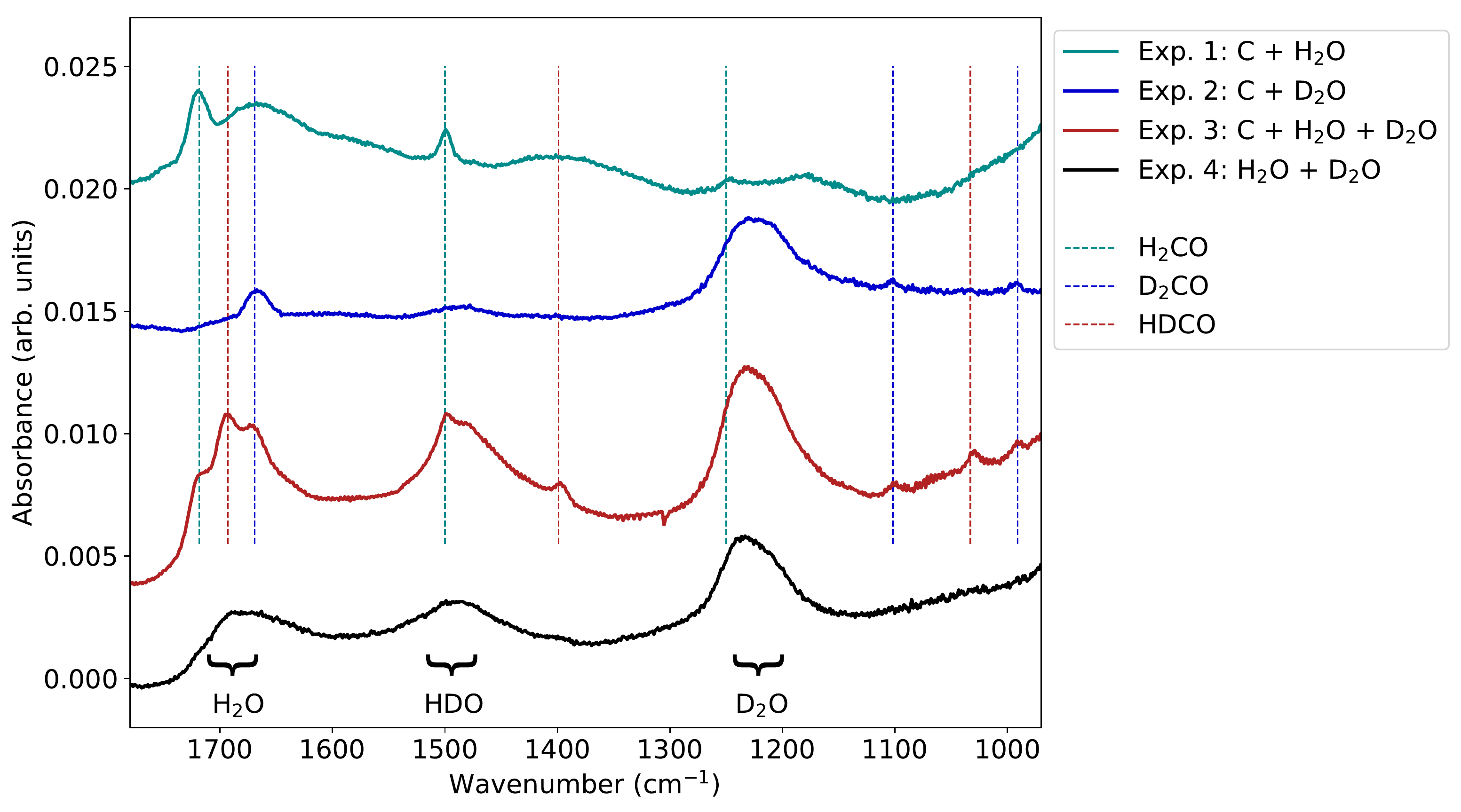}
    \caption{RAIR spectra of the four experiments outlined in \cref{tab:exps}. The vertical lines indicate peak positions of the formaldehyde and water isotopologues.}
    \label{fig:experiments}
\end{figure*}

The RAIR spectra corresponding to experiments $1-4$ are depicted in \cref{fig:experiments} for the wavenumber range specific to the CO stretch, \ce{CH2} scissoring and rocking modes of the three formaldehyde isotopes of interest here: $1750-980$ cm$^{-1}$. The peak positions for all relevant molecules are indicated with a line at the corresponding wavenumber in the figure and are listed in the Supporting Information. Experiments 1 and 2 show clear \ce{H2O} / \ce{H2CO} and \ce{D2O} / \ce{D2CO} detections, respectively. This also points towards the fact that indeed intersystem crossing has occurred on a short timescale. Integrating the CO stretch area under the peak, and taking into account the corresponding absorption band strength \cite{Hidaka:2009, Strickler:1982}, this leads to a \ce{H2CO}/\ce{D2CO} ratio of about 1.2. In other words, we find a very small kinetic isotope effect, that we attribute to the overall isotope effect of the various reactions in the reaction sequence \ce{C + H2O_{ASW} -> H2CO_{ASW}}. Experiment 3 shows that all three formaldehyde isotopologues have been formed simultaneously from a co-deposition of \ce{C + H2O / HDO / D2O}. No kinetic isotope effect can be calculated in experiment 3 because the band strength of the CO stretch mode for HDCO is not reported and the three carbonyl stretch modes overlap with each other and with the \ce{H2O} bending mode. Control experiment 4 indicates that no formaldehyde is formed without impacting C-atoms and also no effect is seen due to potential contaminants in the C-atom source.
These experimental results are fully in line with the computational outcomes presented above in \cref{tab:deuterium}.

We conclude this work emphasizing that we have found an intricate reaction mechanism for the formation of formaldehyde from carbon atoms on top of amorphous solid water, modelled via an ice cluster approach. This work now reconciles the mismatch between gas-phase and solid-state experimental and theoretical efforts through the finding that the water assisted mechanism (\textbf{2}) relies on the collective interplay of the water ice network leading to a concerted proton transfer. Therefore, it operates much more efficiently than the internal hydrogen migration mechanism (\textbf{1}), and explains the lack of detection of reaction products in matrix-isolation experiments.\cite{Schreiner2006,Ortman1990} Furthermore, we expand on previous computational work\cite{Shimonishi2018, Duflot2021} discussing the initial stages of \ce{^3C + H2O} interaction with ASW. 

The low or null activation barriers for the water-assisted mechanism explain the rapid solid-state formation of \ce{H2CO} on amorphous solid water seen experimentally. They are also consistent with the minute kinetic isotope effect observed for the formation of \ce{D2CO}, in contrast with gas-phase studies\cite{Hickson2016} and in agreement with a recent laboratory effort.\cite{potapov2021} The internal isomerization mechanism (Mechanism \textbf{1}) in turn, analogous to the gas-phase process,\cite{Hickson2016, Li2017} presents barriers too high to be relevant under interstellar cold conditions as those found in molecular clouds.


From an astrochemical point-of-view our results serve to highlight the following implications:

\begin{itemize}
    \item The evidence of a catalytic effect of water -- and potentially other hydrogen bonded networks -- suggests that proton transfer reactions may operate in interstellar ices in the presence of highly polarizing adsorbates;
    \item The formation of \ce{H2CO} in early stages of a molecular cloud lifetime points to an early formation of carbon bearing (complex) organic molecules;
    \item The deuterium fractionation of observed formaldehyde in cold regions will be influenced by the reaction route presented here, and should be taken into account along with the deuterium fractionation expected from the main formation route of formaldehyde in later states of a molecular cloud \ce{CO + 2H/D};
    \item The formation route proposed here  opens new avenues for several astrochemical reaction networks, \emph{e.g.}, formation of methanol via subsequent hydrogen additions, and reactions with other radicals that are abundant in the water ice phase, worth to be explored by astrochemical models
\end{itemize}

\begin{acknowledgement}

Ko-Ju Chuang and Melissa McClure are greatly thanked for stimulating discussions. We also thank the anonymous reviewers for their suggestions.
Computer time was granted by the state of Baden-Württemberg through bwHPC and the German Research Foundation (DFG) through grant no. INST 40/467-1FUGG is greatly acknowledged. G.M thanks the Alexander von Humboldt Foundation for a post-doctoral research grant. We thank the Deutsche Forschungsgemeinschaft (DFG, German Research Foundation) for supporting this work by funding EXC 2075 - 390740016 under Germany’s Excellence Strategy. T.L. is grateful for support from the Dutch Research Council (NWO) via a VENI fellowship (grant no. 722.017.008). Astrochemistry in Leiden is supported by the Dutch Astrochemistry Network (DAN) II (Project No. 648.000.029).
G.F. acknowledges financial
support from the Russian Ministry of Science and Higher Education via the State
Assignment Contract. FEUZ-2020-0038. This work has been supported in part by the Danish National Research Foundation through the Center of Excellence `InterCat' (Grant agreement no.: DNRF150)

\end{acknowledgement}

\begin{suppinfo}

Details on the theoretical framework as well as the experimental set-up can be found in the Supporting Information. Furthermore, intrinsic reaction coordinate calculations, consistency tests for the barrierless conversion and transition state geometries are included as well.

\end{suppinfo}

\bibliography{achemso-demo}


\begin{tocentry}

\includegraphics[width=1.0\linewidth]{figures/toc.png}

\end{tocentry}

\end{document}